%% file: article.tex
\documentclass[12pt]{spieman}  
\usepackage{amsmath,amsfonts,amssymb}
\usepackage{graphicx}
\usepackage{setspace}
\usepackage{tocloft}
\usepackage{comment}
\usepackage{color}
\usepackage{soul}

\title{Torque cancellation effect of Intensity noise for Cryogenic sub-Hz cROss torsion bar detector with quantum NOn-demolition Speed meter (CHRONOS)}

\author[a,b,d\dag]{Daiki Tanabe}
\author[c,d,a,b\ddag]{Yuki Inoue}
\author[e,a]{Vivek Kumar}
\author[c,d]{Miftahul Ma'arif}
\author[c,d]{Ta-Chun Yu}

\affil[a]{Institute of Physics, Academia Sinica, Nangang, Taipei, 015011, Taiwan}
\affil[b]{Institute of Particle and Nuclear Studies (IPNS), High Energy Accelerator Research Organization (KEK), Tsukuba, Ibaraki 305-0801, Japan}
\affil[c]{Physics Department, National Central University, Taoyuan 32001, Taiwan}
\affil[d]{Center for High Energy and High Field Physics, National Central University, Taoyuan 32001, Taiwan}
\affil[e]{Department of Physics, Institute of Applied Sciences and Humanities, GLA University, Mathura 281406, India}

\cftpagenumbersoff{figure}
\cftpagenumbersoff{table} 

\begin{document} 
\maketitle

\begin{abstract}
Detection of sub-Hz gravitational waves is of significant importance in various aspects of astrophysics. It enables us to detect intermediate-mass black-hole mergers, issue early alerts for gravitational-wave events, and explore a landscape of the stochastic background. Cryogenic sub-Hz cROss torsion bar detector with quantum NOn-demolition Speed meter (CHRONOS) is a proposed gravitational-wave detector based on a speed meter that uses torsion bars as end test masses. It aims to achieve $3\times 10^{-18} \, {\rm Hz}^{-1/2}$ in strain sensitivity at 1~Hz with its prototype design, to detect $\mathcal{O}(10^4)M_\odot$ intermediate-mass black-hole mergers at 100 Mpc with a signal-to-noise ratio of 3. We point out that the torsion-bar-based speed meter can suppress noise originating from laser intensity fluctuation by cancellation of torque on the bar and by a balanced homodyne readout. We propose, for the first time, an intensity-noise formula for a gravitational-wave detector employing a Sagnac speed-meter configuration with a torsion bar. Based on our intensity-noise model, we evaluate it for a 2.5-m-arm prototype of CHRONOS. The projected intensity noise was $2.9 \times 10^{-20} \, {\rm Hz}^{-1/2}$ which was sufficiently low to detect a binary intermediate-mass black-hole merger.
\end{abstract}



{\noindent \footnotesize\textbf{\dag}Daiki Tanabe,  \linkable{tana2431.ts@gmail.com} }
{\noindent \footnotesize\textbf{\ddag}Corresponding author: Yuki Inoue,  \linkable{iyuki@ncu.edu.tw} }

\begin{spacing}{1}

\input{1_intro}

\input{2_intensity}
\input{3_offset}
\input{4_evaluation}

\input{5_discussion}

\input{6_conclusion}

\acknowledgments     
We thank Masashi Hazumi for his academic advice during the preparation of this manuscript. We appreciate Rick Savage and Sadakazu Haino for discussion on the evaluation method of bulk deformation. Chao Shiuh helped our mechanical simulation. We are grateful to Tsung-Chieh Ho, Ko-Han Chen, Aloysius Niko, and Cheng-Han Chan, who contributed to develop the input optics of CHRONOS. Avani Patel, Afif Ismail, and Henry Tsz-King Wong provided fruitful discussions on the sensitivity calculation of CHRONOS. Y.I. acknowledges support from NSTC, CHiP, and Academia Sinica in Taiwan under Grant No.114-2112-M-008-006- and No.AS-TP-112-M01.

\bibliography{reference}   
\bibliographystyle{chronosbib} 

\end{spacing}
\end{document}

%% file: 1_intro.tex
\section{Introduction}
\label{sect:intro_section}

With the continuing enhancement of gravitational-wave (GW) detector sensitivity, statistical investigations of GW populations—most prominently binary black holes and neutron stars—are becoming increasingly feasible. The initial phase of the fourth joint observing run (O4) of LIGO, Virgo, and KAGRA, which concluded in 2024, reported 128 additional candidate events from compact binary coalescences (CBCs)~\cite{gwtc4}. In parallel, major international initiatives, including the planned Laser Interferometer Space Antenna (LISA), are striving to open new observational windows in the millihertz regime, thereby extending the frequency coverage of GW astronomy and providing complementary access to earlier cosmic epochs.

The frequency of GWs emitted during the inspiral phase of CBCs scales inversely with the total mass of the system, such that different frequency bands probe distinct astrophysical populations. Frequencies below 1 Hz, in particular, correspond to the characteristic inspiral signals of binaries hosting intermediate-mass black holes (IMBHs) with component masses in the range $10^3$–$10^5\,M_\odot$~\cite{IMBH_calc}. IMBHs are hypothesized to represent the missing evolutionary stage between the well-established stellar-mass black holes and the supermassive black holes observed in galactic nuclei. Achieving sensitivity in the sub-hertz frequency band is therefore a key to resolving this long-standing mystery of black-hole formation and growth. Although several binary black hole mergers with component masses of $\mathcal{O}(10^2)\,M_\odot$ have been detected, robust observational evidence for binaries above $10^3\,M_\odot$ remains elusive~\cite{IMBH_GW190521_1,IMBH_GW231123}.

A Sagnac speed meter based on a cryogenic torsion bar, named the Cryogenic sub-Hz cROss torsion bar detector with quantum NOn-demolition Speed meter (CHRONOS), has been proposed as a novel gravitational-wave (GW) detector optimized for the sub-hertz band~\cite{CHRONOS}. The concept of using torsion bars as end test masses in GW detectors has been pioneered by the Torsion-Bar Antenna (TOBA) and the Torsion Pendulum Dual Oscillator (TorPeDO)~\cite{TOBA,Torpedo}. A key advantage of the torsion-bar design lies in the yaw rotation mode of a fiber-suspended bar, whose resonant frequency is significantly lower than that of the translational pendulum mode of a suspended mirror, thereby providing strong suppression of seismic noise.  
In parallel, the speed-meter principle reduces the quantum noise scaling in frequency space from $1/f^2$ to $1/f$~\cite{speedmeter}. Combining these two features---the torsion bar and the speed-meter topology---enables a compact detector design while maintaining sufficient low-frequency sensitivity.  
The CHRONOS program envisions three configurations with different arm lengths, adapted to existing GW test facilities: 2.5 m, 40 m, and 300 m. In its ultimate 300-m configuration, CHRONOS aims to reach a strain sensitivity of $1\times 10^{-18}\,{\rm Hz}^{-1/2}$ at 1 Hz, sufficient to detect gravitational waves from $\mathcal{O}(10^4)\,M_\odot$ binary IMBH mergers out to a luminosity distance of $\sim 380$ Mpc with a signal-to-noise ratio (SNR) of 3. The intermediate 40-m configuration targets $2\times 10^{-18}\,{\rm Hz}^{-1/2}$, while the 2.5-m prototype aims for $3\times 10^{-18}\,{\rm Hz}^{-1/2}$. Even in this prototype setup, CHRONOS would retain the capability to detect $\mathcal{O}(10^4)\,M_\odot$ binary IMBH mergers within $\sim 100$ Mpc with an SNR of 3.

In this study, we present the first attempt to construct an intensity-noise model for a torsion-bar detector combined with a speed-meter topology. We demonstrate that a common-mode cancellation mechanism against laser intensity noise operates in this configuration by combining numerical analysis with the finite-element analysis (FEA) software COMSOL and an analytical model~\cite{COMSOL}. Specifically, (i) we take into account the torque cancellation effect arising from the geometrical symmetry of mirrors mounted on the torsion bar, and (ii) we calculate the noise using the transfer function of CHRONOS operating as a speed meter obtained with FINESSE3~\cite{finesse_software}. In this way, we incorporate a concrete optical system into the simulation and evaluate the intensity-noise suppression effect inherent to the torsion-bar speed-meter configuration.

In the following part of this paper, we provide a formula of intensity noise of a torsion-bar-based speed meter and evaluate it based on 2.5-m configuration of CHRONOS. Sec.~\ref{sect:intensity_section} describes a mathematical model of intensity noise in terms of spacetime strain. Sec.~\ref{sect:offset_section} presents simulations of the surface displacement of the torsion bar with a realistic beam-position misalignment as a part of the intensity noise model. Sec.~\ref{sect:evaluation_section} evaluates the intensity noise of CHRONOS using the designed parameters and the laser-intensity fluctuation measured in a previous study. Sec.~\ref{sect:discussion_section} discusses case studies of 40-m and 300-m options, followed by an interpretation of the simulation results.

%% file: 2_intensity.tex
\section{Intensity-noise model\label{sect:intensity_section}}

\subsection{General detection principle of speed meter with torsion bar}

\begin{figure}
\begin{center}
\begin{tabular}{c}
\includegraphics[width=12.0cm]{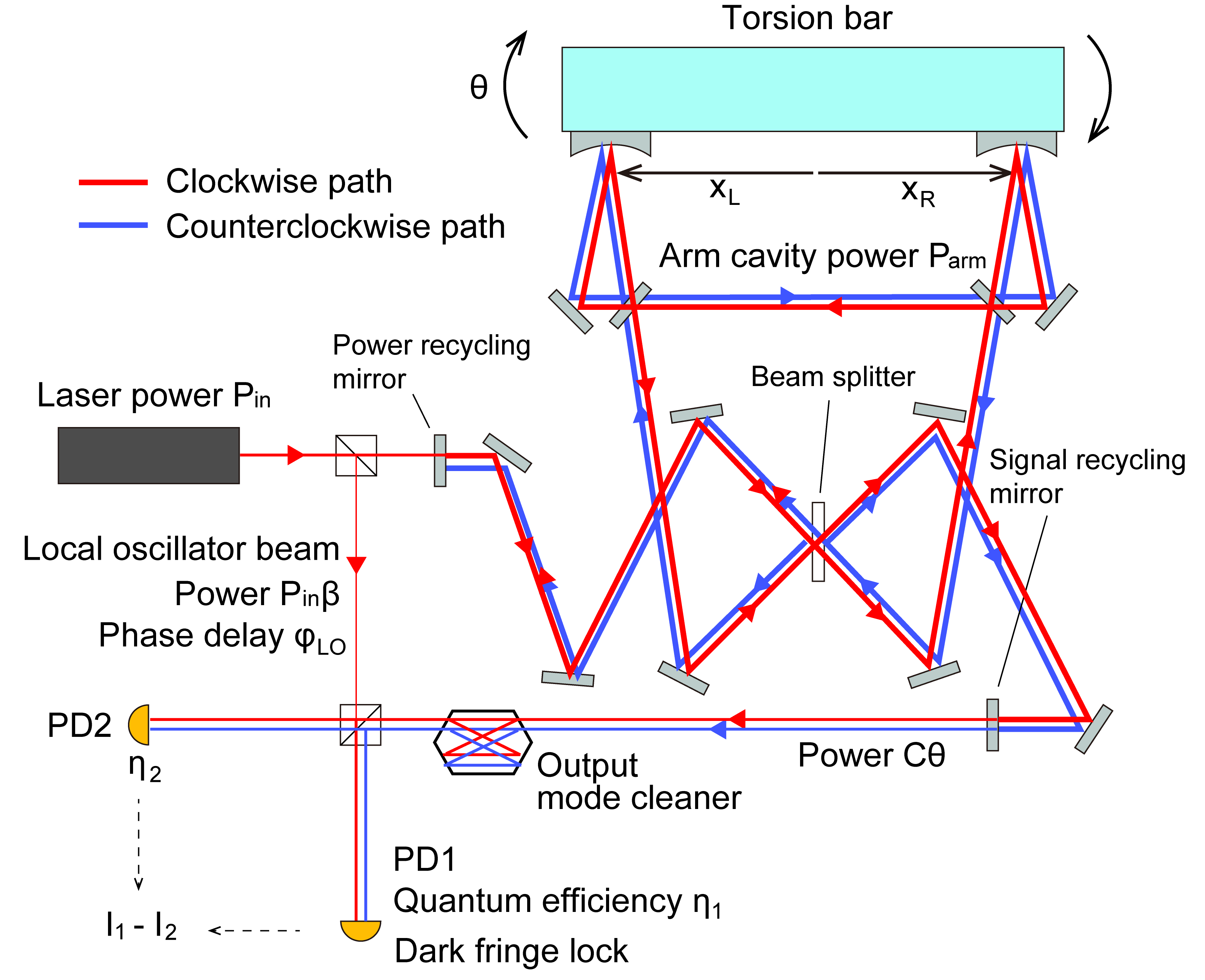}
\end{tabular}
\end{center}
\caption[]
{ \label{fig:chronos_optics_simplified}
Simplified optical path of CHRONOS. One torsion bar and balanced-homodyne detection part are shown.
}
\end{figure}

A simplified schematic view of CHRONOS optics is shown in Fig.~\ref{fig:chronos_optics_simplified}. Here we focus only on a single bar because of the symmetries of the two bars and their associated optical layouts. An input laser beam is split into clockwise and counterclockwise paths by the first beam splitter. The beam on each path hits the left or right mirror attached to the bar. The mirrors are attached at horizontally symmetric positions on the bar and form triangular cavities at each side of the bar. The phase difference between the two paths accumulated during resonance in the triangular cavities is proportional to the rotation velocity of the bar, while GW strain generates a rotation angle. As the rotation velocity and angle are related by $\dot{\theta}(\omega)=i\omega\theta(\omega)$, this configuration can reduce the slope of the quantum noise spectrum from $1/f^2$ to $1/f$, which makes it particularly advantageous at low frequencies\cite{speedmeter}. 

Relationship between the rotation angle and GW strain follows a known second-order differential equation relating angle and torque. A torsion bar is specifically sensitive to the cross-mode GW denoted by $h_{\times}(\omega)$. The bar response is~\cite{shimoda_phd}
\begin{equation}
\label{eq:theta_GW}
\theta_{\rm GW}(\omega)=\frac{1}{2}\frac{\omega^2}{\omega^2-i\frac{\omega_0\omega}{Q}-\omega_0^2}h_{\times}(\omega) \simeq \frac{h_{\times}(\omega)}{2},
\end{equation}
where $\omega_0$ is the resonant frequency of the yaw rotation mode and $Q$ is a quality factor. We neglected the $\omega_0/Q$ term because the $Q$ of sapphire, typically on the order of $10^8$, is sufficiently larger than $\omega_0$. Since $\omega_0$ lies in the mHz region with our suspension design, it can be neglected compared with $\omega$.

Since CHRONOS has two bars aligned in a cross shape, we have to consider the differential rotation angle and the antenna pattern. Assuming the bars are initially at rest, the net rotation of the two bars is simply twice that of one bar because cross-mode GW rotates them in opposite directions~\cite{TOBA_calc}. In an actual observation site, plus-mode and cross-mode are mixed in a finite ratio depending on the location and the direction of the detector. Here we assume a uniform antenna pattern for simplification. Then, the expectation value of the rotation angle is~\cite{antenna_pattern}
\begin{equation}
\label{eq:antenna}
\begin{array}{lll}
\left< \theta_{\rm GW,2bar}(\omega) \right>&=&\left[ \int_0^1 (1+\lambda^2) \left| F(j,k,l) \right|^2 d\lambda \right]^{\frac{1}{2}}2\theta_{\rm GW} \\
&\simeq& \frac{2}{\sqrt{3}}h_\times (\omega),
\end{array}
\end{equation}
where $j$, $k$, $l$ are angles between the detector-bisector direction and the local meridian, the latitude the detector, and the direction of the incoming GW, respectively. The $F(j,k,l)$ represents the antenna pattern which is assumed to be unity here. The $\lambda$ is a parameter related to the GW source distribution. By inverting Eq.~(\ref{eq:antenna}) and substituting the measured differential rotation angle we reconstruct $h_\times (\omega)$ and its spectrum.

The differential phase, corresponding to the differential rotation angle, is detected by a balanced-homodyne configuration, which is required for a speed meter to transcend the standard quantum limit by tuning the homodyne detection angle independently of the GW phase\cite{speedmeter,speedmeter_homodyne}. The merged beam from the two paths is split again at the output port. Simultaneously, a local-oscillator beam of the same wavelength as the input laser is injected to decouple the common mode and the differential mode. The differential mode remains after taking the difference of the currents from the two PDs. We reconstruct GW strain from the differential current by applying an inverse filter of the sensing function, which models the transfer function from rotation angle to differential power.

The torsion bar geometry and the balanced-homodyne detection make its intensity-noise model distinct from thet of a conventional Fabry--P$\acute{e}$rot Michelson interferometer. The intensity-noise model of Advanced LIGO includes an offset of differential arm length (DARM), an imbalance of two arm cavities, a sideband transmittance through an output mode cleaner (OMC), and radiation pressure on the mirrors~\cite{ligo_intensity_model_part2, ligo_intensity_model_part3}. Among these four components, we focus only on the radiation pressure terms for CHRONOS because the other three components are common and subdominant. All of them are smaller than the radiation pressure terms by a factor of $10^{-4}$ or less in the case of Advanced LIGO. The effect of radiation pressure should be specifically modeled for a torsion bar with consideration of torque cancellation. Intensity noise of the local-oscillator beam at the balanced-homodyne-detection stage, including an imbalance of the beam splitter's splitting ratio and a mismatch of the PDs' quantum efficiency, is another new effect that should be considered for a speed meter.

\subsection{Photon pressure mismatch on the torsion bar}
A laser beam exerts photon pressure, which applies a force onto the surface of the torsion bar. The power from the input optics is amplified by arm cavities and recycling cavities before hitting the bar. The amplification gain is mainly determined by cavity finesse and is also affected by detuning configuration. When the effective DC offset power at the mirror is $P_{\rm arm}$ and the incident angle is $\varphi$, the force applied by that beam in the short axis direction of the bar is $P_{\rm arm}\cos\varphi_{\rm i}/c$, using the speed of light $c$~\cite{ligo_pcal}. 

We write the bar's transfer function from force to displacement along the beam axis as $H(x,y;\omega)$ using horizontal and vertical coordinates $(x, y)$ on the bar surface. Here the origin of $(x,y)$ is the center of mass of the bar. The rotation angle induced by the photon pressure is
\begin{equation}
\label{eq:angle}
\theta_{\rm i}(\omega)=\frac{2P_{\rm arm}\cos\varphi_{\rm i}}{c}\frac{H(x_{\rm i},y_{\rm i};\omega)}{x_{\rm i}}, \quad \left( {\rm i}={\rm L,R} \right).
\end{equation}

The net rotation angle is
\begin{equation}
\label{eq:theta_bar_raw}
\begin{array}{lll}
\theta_{\rm PP}(\omega)&\equiv& \frac{\theta_{\rm L}(\omega)-\theta_{\rm R}(\omega)}{2} \\
&\simeq& \frac{P_{\rm arm}}{c}\left( \frac{H(x_{\rm L},y_{\rm L};\omega)}{x_{\rm L}}-\frac{H(x_{\rm R},y_{\rm R},\omega)}{x_{\rm R}} \right).
\end{array}
\end{equation}
Here we neglected square terms of $\varphi_{\rm i}$. When we define a horizontal beam position mismatch factor $\alpha$ as
\begin{equation}
\label{eq:alpha}
\alpha\equiv \frac{x_{\rm L}-x_{\rm R}}{\bar{x}}, \quad \bar{x}\equiv \frac{x_{\rm L}+x_{\rm R}}{2},
\end{equation}
then the Eq.~(\ref{eq:theta_bar_raw}) can be written as
\begin{equation}
\label{eq:theta_bar_raw_alpha}
\begin{array}{lll}
\theta_{\rm PP}(\omega)&\simeq& \frac{P_{\rm arm}}{c}\frac{\Delta H(\omega)-\bar{H}(\omega)\alpha}{\bar{x}},
\end{array}
\end{equation}
where
\begin{eqnarray}
\label{eq:H_delta_bar}
\Delta H(\omega)\equiv H(x_{\rm L},y_{\rm L};\omega)-H(x_{\rm R},y_{\rm R};\omega),\\
\bar{H}(\omega)\equiv \frac{H(x_{\rm L},y_{\rm L};\omega)+H(x_{\rm R},y_{\rm R};\omega)}{2}.    
\end{eqnarray}
Here we neglected square terms of $\alpha$.

For a finite rigidity material, $H(x,y;\omega)$ is a combination of translational motion, rotation, and bulk deformation, which makes it complicated to model\cite{sudarshan_phd,bulk_document}. We model the precise shape of $H(\omega;x,y)$ by finite-element analysis (FEA) in Sec.~\ref{sect:evaluation_section}. However, its behavior at low frequency is dominated by yaw rotation. Replacing the torque of GW strain $I\omega^2 h_\times (\omega)/2$ in Eq.~(\ref{eq:theta_GW}) with the torque of the photon pressure, the yaw rotation angle of a rigid torsion bar is
\begin{equation}
\label{eq:rotation_model}
\theta_{\rm i}^\prime(\omega)\simeq \frac{1}{I\omega^2}\frac{2P_{\rm arm}\cos\varphi_{\rm i}}{c}x_{\rm i},
\end{equation}
where $I$ is the moment of inertia of the bar around the axis of rotation.

Comparing Eq.~(\ref{eq:angle}) and Eq.~(\ref{eq:rotation_model}), the explicit form of $H(x_{\rm i},y_{\rm i};\omega)$ is approximated to be
\begin{equation}
\label{eq:H_TF}
H^\prime(x_{\rm i},y_{\rm i};\omega)\equiv \frac{x_{\rm i}^2}{I\omega^2},
\end{equation}
at frequencies sufficiently higher than the yaw resonant mode. We confirm this functional form in Sec.~\ref{sect:offset_section}. 

Intensity noise is characterized by relative intensity noise (RIN), which is a ratio of the intensity-noise spectrum to the DC offset of the intensity. Therefore, the power injected to the mirror is replaced by $P_{\rm arm}(1+{\rm RIN}(\omega))$. We can neglect the offset term $P_{\rm arm}$ because the DC power is reflected back by the power recycling mirror and only the frequency-dependent component is transmitted to the output port~\cite{ando_phd}. Since the RIN is classical fluctuation, we can also assume that the RINs of counterclockwise and clockwise paths are sufficiently coherent in the frequency range below kHz, where the noise wavelength is much longer than the error of optical path length. Then, the net rotation angle becomes
\begin{equation}
\label{eq:theta_bar}
\theta_{\rm PP}(\omega)\simeq \frac{P_{\rm arm}{\rm RIN(\omega)}}{c}\frac{\Delta H(\omega)-\bar{H}(\omega)\alpha}{\bar{x}}.
\end{equation}
The Eq.~(\ref{eq:theta_bar}) implies that the RIN is suppressed by a factor $\alpha$. 

The differential rotation angle is detected through a sensing function in a closed-loop feedback control. As in conventional Michelson-type GW detectors, the CHRONOS interferometer is locked by digital feedback control of the test mass. When we write the {\it sensing function} and the {\it open-loop transfer function} as $C(\omega)$ and $G(\omega)$, respectively, the {\it closed-loop transfer function} is expressed as $C(\omega)/(1+G(\omega))$~\cite{ligo_calibration,virgo_calibration,kagra_calibration}. Here the $G(\omega)$ depends on the digital filter design for feedback. However, it typically drops to values smaller than 1 above the unity gain frequency. The target frequency of CHRONOS requires us to set the unity gain frequency below the sub-Hz range so as not to apply feedback to low frequency GW signals. Hence, we can neglect $G(\omega)$ in most frequency regions. In this scheme, The differential power generated by the differential angle is simply written as
\begin{equation}
\label{eq:sensing}
\Delta P_{\rm bar}(\omega)=C(\omega)\theta_{\rm PP}(\omega),
\end{equation}
which is converted into current at the balance homodyne detection port.

\subsection{Fluctuation of the local oscillator beam}
The beam from the torsion bar is merged with the local-oscillator beam at the beam splitter. It enables us to demodulate the signal at the laser frequency and decouple the common mode and the differential mode using two beam paths. The local-oscillator beam is sampled from the input laser and attenuated by a factor $\beta$. The local-oscillator beam shares the intensity noise with the main laser. 

The sum of the electric fields of the beams from the bar, $E_{\rm bar}$, and from the local oscillator generates common-mode and differential-mode components. Power at each PD is
\begin{equation}
\label{eq:field}
\begin{array}{lll}
P_{\rm PD1}(\omega)&=&\left| E_{\rm bar}(\omega)+iE_{\rm LO}(\omega) \right|^2, \\
P_{\rm PD2}(\omega)&=&\left| iE_{\rm bar}(\omega)+E_{\rm LO}(\omega) \right|^2,
\end{array}
\end{equation}
where $E_{\rm bar}(\omega)$ and $E_{\rm LO}(\omega)$ are the electric fields from the torsion bar and the local oscillator in each path, respectively.

The differential-mode terms are given by Eq.~(\ref{eq:sensing}). On the other hand, although the common mode is canceled by taking the differential of the two PDs, it can remain in practice due to a mismatch of the PDs' quantum efficiency. Therefore, the intensity fluctuation on the local-oscillator beam has to be accounted as intensity-noise component. The $\left| E_{\rm bar}(\omega) \right|^2$ term can be neglected because it becomes proportional to a product of the DARM offset and the mismatch factor of the balanced-homodyne readout.

Power of the local-oscillator beam, here we denoted by $\left| E_{\rm LO}(\omega) \right|^2$, is explicitly written as 
\begin{equation}
P_{\rm LO}(\omega)= P_{\rm in}(1+{\rm RIN}(\omega))\beta.
\end{equation}
Here we can neglect the DC term when we operate the interferometer with a dark fringe lock. The PDs convert it to current as
\begin{equation}
\begin{array}{lll}
I_{\rm PD1}&=&P_{\rm in}{\rm RIN}(\omega)\beta\eta_{\rm PD1},\\
I_{\rm PD2}&=&P_{\rm in}{\rm RIN}(\omega)\beta\eta_{\rm PD2},
\end{array}    
\end{equation}
where $\eta_{\rm {PD1}}$ and $\eta_{\rm {PD2}}$ represent the total efficiency from power to current, including the beam splitting ratio and the PDs' quantum efficiency. Here we define a homodyne mismatch parameter,
\begin{equation}
\gamma\equiv \frac{\eta_{\rm PD1}-\eta_{\rm PD2}}{\bar{\eta}}, \quad \bar{\eta}\equiv \frac{\eta_{\rm PD1}+\eta_{\rm PD2}}{2}
\end{equation}
Then, the differential current is
\begin{equation}
\label{eq:I_LO}
\Delta I_{\rm LO}= I_{\rm PD1}-I_{\rm PD2}=P_{\rm in}{\rm RIN}(\omega)\beta\bar{\eta}\gamma.
\end{equation}

We also neglected the $\left| E_{\rm bar}(\omega) \right|^2$ term in Eq.~(\ref{eq:field}) because its DC term is minimized by dark fringe lock, and its frequency-dependent term contains a product of $\alpha$ and $\gamma$, which makes it negligibly small. The coupling of $\alpha$ and $\gamma$ in the differential mode was neglected as well.

\subsection{Reconstruction of spacetime strain noise}
Based on Eq.~(\ref{eq:theta_bar}), Eq.~(\ref{eq:sensing}), and Eq.~(\ref{eq:I_LO}), the intensity-noise coupling components in the differential current are written as
\begin{equation}
\label{eq:current_spectrum}
\begin{array}{l}
S_I(\omega)=\sqrt{2\left| \Delta P_{\rm bar}(\omega)\bar{\eta} \right|^2+\left| \Delta I_{\rm LO}(\omega) \right|^2}\\
={\rm RIN}(\omega) \bar{\eta} \sqrt{2\left| C(\omega) \frac{P_{\rm arm}}{c}\frac{\Delta H(\omega)-\bar{H}(\omega)\alpha}{\bar{x}} \right|^2+\left| P_{\rm in}\beta\gamma \right|^2}
\end{array}
\end{equation}
in terms of noise spectrum. Here we neglected the coupling of $\Delta P_{\rm bar}$ and $\gamma$ because it is a second-order term of fluctuation. A noise spectrum of rotation angle is reconstructed from the current by the inverse sensing function and the PD efficiency as
\begin{equation}
\label{eq:theta_measured_spectrum}
S_{\theta}(\omega)=C_{\rm modeled}^{-1}(\omega) \bar{\eta}^{-1} S_I(\omega).
\end{equation}
We regard the modeled $C(\omega)$ as identical to the true transfer function and treat their discrepancy as a systematic error. Substituting (\ref{eq:current_spectrum}) into Eq.~(\ref{eq:theta_measured_spectrum}) and considering the antenna pattern defined in Eq.~(\ref{eq:antenna}), the noise spectrum in terms of spacetime strain is explicitly written as 
\begin{equation}
\label{eq:h_total_spectrum}
\begin{array}{l}
S_h(\omega)\simeq {\rm RIN}(\omega) \\
\cdot \sqrt{ \frac{3P_{\rm arm}^2} {2c^2\bar{x}^2} \left| \Delta H(\omega)-\bar{H}(\omega)\alpha \right|^2+\left| C_{\rm modeled}^{-1}(\omega) \right|^2 P_{\rm in}^2 \beta^2\gamma^2}.
\end{array}
\end{equation}

To evaluate the sensitivity with this function, we must determine
(i) the common-mode response \(H\) and the differential-mode response \(\Delta H\),
(ii) a representative beam-mismatch parameter \(\alpha\), and
(iii) the sensing function \(C\) for the assumed speed-meter configuration.
These quantities are defined and derived in Sec.~3. 

%% file: 3_offset.tex
\section{Transfer function of bar surface displacement\label{sect:offset_section}}

In this study, we evaluate the back-action induced by laser power 
fluctuations on the torsion-bar test mass, explicitly accounting for 
its internal deformation modes. To this end, three key ingredients are addressed:  

\begin{itemize}
  \item[(i)] \textbf{Common- and differential-mode responses}:  
  We refined the transfer function \(H(x,y;\omega)\) of the CHRONOS torsion bar---which had previously been approximated as a simple \(1/f^2\) dependence in Eq.~(\ref{eq:H_TF})---through finite-element analysis (FEA). 
  This refinement allows us to describe both the common-mode response \(H\) and the differential-mode response \(\Delta H\) in a frequency-dependent manner.  

  \item[(ii)] \textbf{Beam-mismatch parameter \(\alpha\)}:  
  The dependence of the refined transfer function on the representative beam-mismatch parameter \(\alpha\) was explicitly investigated to quantify the impact of imperfect mode overlap on the induced back-action.  

  \item[(iii)] \textbf{Sensing function \(C\) of the speed-meter}:  
  While the present section focuses on the mechanical response and mismatch effects, the sensing function \(C(\omega)\) of the assumed speed-meter readout was introduced to combine with (i) and (ii) for the full sensitivity evaluation.  
\end{itemize}

The simulations for (i) and (ii) were performed using COMSOL Multiphysics 5.4 (hereafter COMSOL)~\cite{COMSOL}. 
Previous comparative studies have demonstrated that COMSOL and ANSYS yield consistent results for bulk deformation of test masses under millimeter-scale meshing and nanonewton-scale external forces, and that this FEA-based transfer-function evaluation method has been experimentally validated with photon calibrators in both LIGO and KAGRA~\cite{kagra_pcal,bulk_document}.  

In the present work, we adopted sapphire material parameters identical to those of the KAGRA test masses, as summarized in Tab.~\ref{tab:sapphire_parameters}~\cite{O3GK}. 
The mesh size was chosen to be finer than 1~mm in the laser beam cross-section region and finer than 1.8~cm in the remaining regions.  

\subsection{Common and Differential Torsion bar response: $\bar{H}(\omega)$ and $\Delta H(\omega)$\label{subsec:H}}
\begin{table}[tbp]
\centering
    \begin{tabular}{cc}
    \hline
    Parameter (unit) & Value \\
    \hline \hline
    Density (${\rm kg/m^3}$) & $4 \times 10^3$ \\
    Young's modulus (GPa) & $4 \times 10^2$ \\
    Poisson's ratio & $0.29$ \\
    \hline
    \end{tabular}
\caption{Mechanical parameters of sapphire used for FEA simulation~\cite{O3GK}.}
\label{tab:sapphire_parameters}
\end{table}

\begin{figure}[tp]
\begin{center}
\begin{tabular}{c}
\includegraphics[width=12.0cm]{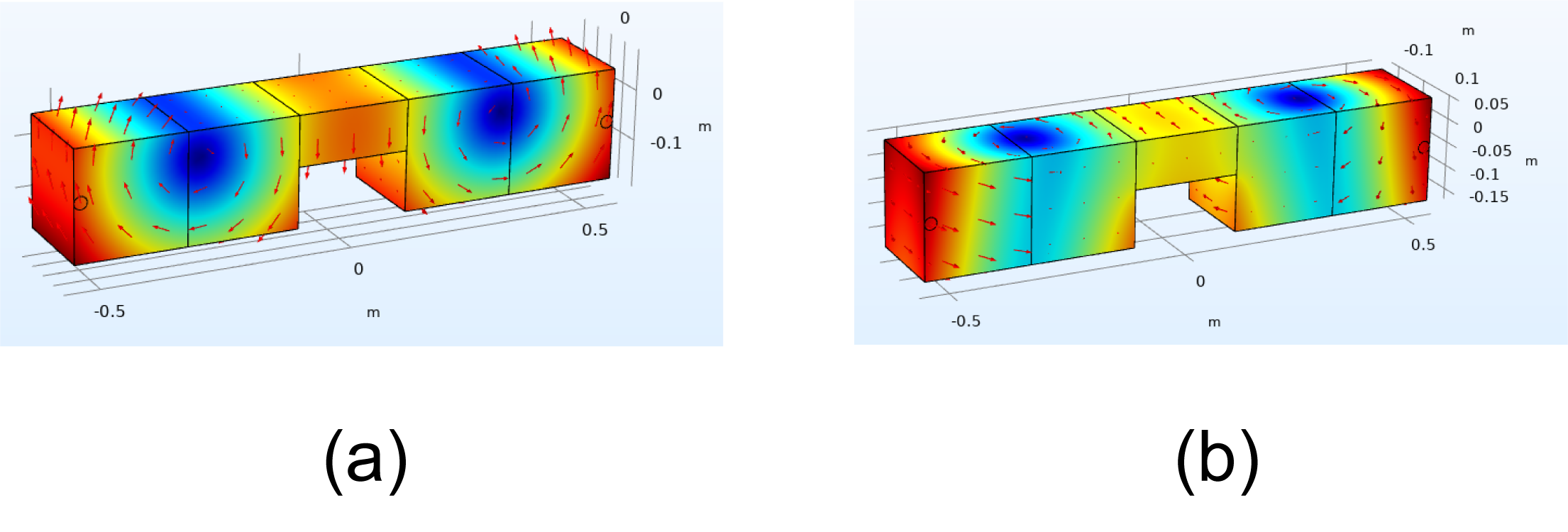}
\end{tabular}
\end{center}
\caption[]
{ \label{fig:bar_modeshape}
Mode shapes of the first and second bulk deformation modes of CHRONOS torsion bar calculated by COMSOL. Arrows indicate direction of surface displacement at each area. The circles of black line at left and right sides represent mirrors. (a) Vertical bending mode at 667 Hz. (b) Horizontal bending mode at 1238 Hz.
}
\end{figure}

First, we performed modal analysis to investigate the resonant frequencies associated with bulk deformation modes of the torsion bar. The obtained eigenfrequencies are summarized in Tab.~\ref{tab:modal}. The lowest mode appears at 667~Hz, followed by the second-lowest mode at 1238~Hz, and their mode shapes are shown in Fig.~\ref{fig:bar_modeshape}. These frequencies are much higher than the yaw rotational resonance, which is expected to lie in the mHz region based on the wire length and bar design. Moreover, the surface displacement of the lowest mode is oriented perpendicular to the beam axis and is therefore not expected to contribute to the sensing. Consequently, in the frequency band of interest between 0.1~Hz and 100~Hz, the response is dominated by rotational and translational motions.

\begin{table}[tbp]
\centering
    \begin{tabular}{cc}
    \hline
    Description of mode shape & Eigenfrequency (Hz) \\
    \hline \hline
    Vertical bending & 667 \\
    Horizontal bending & 1238 \\
    Twisting & 1399 \\
    2nd vertical bending & 2515 \\
    2nd horizontal bending & 3136 \\
    \hline
    \end{tabular}
\caption{First five resonant frequencies of CHRONOS torsion bar's bulk deformation modes.}
\label{tab:modal}
\end{table}

Second, Subsequently, we simulated the surface displacement induced by external forces. Specifically, we applied periodic forces corresponding to the photon pressure of a 1~W laser beam at the designed beam positions. The beams were assumed to impinge on the centers of mirrors attached to both ends of the 110-cm bar. These mirrors were assumed to be chemically, thus rigidly, bonded to the bar using hydro-catalysis bonding. To suppress unwanted pitch rotation, the vertical height of the beams was aligned with the center of mass of the bar. The frequency was swept from $10^{-4}$~Hz to $10^{4}$~Hz in order to obtain the harmonic responses. The average surface displacement within the beam profile provides the main contribution to the detected rotational signal. Using the surface displacement $D(\omega;x_{\rm i},y_{\rm i})$ at each beam obtained by COMSOL simulations, the averaged displacement at each beam was calculated as~\cite{sudarshan_phd,bulk_document}
\begin{equation}
d_{\rm i}=\frac{\iint_{\rm beam, i} dx\, dy\, G(x-a_{\rm i},y-b_{\rm i}) D(x,y,a_{\rm i},b_{\rm i};\omega)}{\iint_{\rm beam, i} dx\, dy\, G(x-a_{\rm i},y-b_{\rm i})},
\end{equation}
where $G(x-a_{\rm i},y-b_{\rm i})$ denotes the Gaussian beam profile centered at $(x,y)=(a_{\rm i},b_{\rm i})$ on the bar surface.

Here, $G(x-a_{\rm i},y-b_{\rm i})$ represents the Gaussian beam profile incident at $(x,y)=(a_{\rm i},b_{\rm i})$ on the bar surface. The beam radii were set according to each design option of CHRONOS, and the integration domain was chosen as a circular area with the corresponding radius, as summarized in Tab.~\ref{tab:beam_radius}. The mirrors were assumed to have a diameter of several inches and be mounted at both ends of the bar. By normalizing each $d_{\rm i}$ with the corresponding beam force, we obtained the transfer function $H(x_{\rm i},y_{\rm i};\omega)$ defined in Eq.~(\ref{eq:angle}).

\begin{table}[tbp]
\centering
    \begin{tabular}{cccc}
    \hline
    Parameter (Unit) & 2.5 m & 40 m & 300 m\\
    \hline \hline
    Beam radius (mm) & 2 & 30 & 70 \\
    \begin{tabular}{c} Beam position\\from the end of the bar (inch) \end{tabular} & 0.5 & 1.5 & 3 \\
    \hline
    \end{tabular}
\caption{Assumed parameters of the main beam geometry~\cite{CHRONOS}.}
\label{tab:beam_radius}
\end{table}

Next, we calculated $\bar{H}(\omega)$ and $\Delta H(\omega)$, as defined in Eq.~(\ref{eq:H_delta_bar}), based on the simulation results. Here, $\bar{H}(\omega)$ was obtained as the average response when either the left or the right beam was injected individually, representing the \textit{common-mode} response. In contrast, $\Delta H(\omega)$ was defined as the difference in response when both beams were simultaneously injected, corresponding to the differential-mode response.

\begin{figure}
\begin{center}
\begin{tabular}{c}
\includegraphics[width=12.0cm]{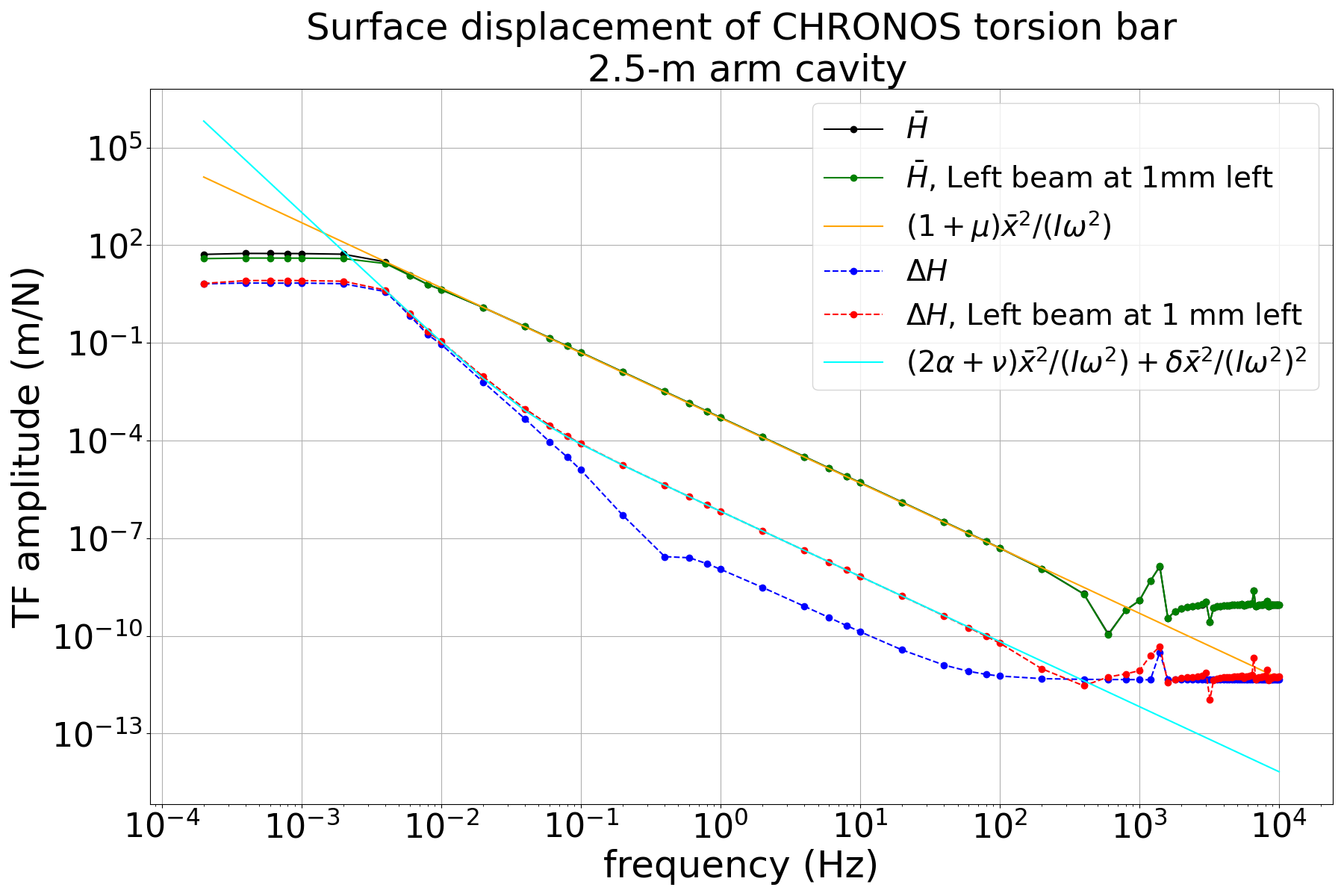}
\end{tabular}
\end{center}
\caption[]
{ \label{fig:twobeams_offset}
Transfer functions from photon force to surface displacement at the mirrors. The CHRONOS 2.5-m option was assumed. Simulated by COMSOL. ({\it Solid black}) Mean value of left or right beam injection at the designed position; ({\it Solid green}) Mean value of left or right beam injection. The left beam was at 1 mm outward while the right beam was at the designed position; ({\it Solid orange}) Model function given in Eq.~(\ref{eq:H_TF}). The moment of inertia was $I=19.9 \ {\rm kg\cdot m^2}$, the mean horizontal position of the beams was $\bar{x}=0.5373 \ {\rm m}$, and the residual parameter was $\mu=0.35$; ({\it Dashed blue}) Differential of left and right mirrors when two beams are simultaneously injected at the respective designed positions; ({\it Dashed red}) Differential when the left beam shifted for 1 mm outward; ({\it Solid cyan}) Model function given in Eq.~(\ref{eq:Delta_H_TF}). The beam position mismatch was $a=0.0021$, $1/f^2$ residual parameter was $\nu=-0.0019$, and the $1/f^4$ magnitude parameter was $\delta=0.0022$.
}
\end{figure}

Figure~\ref{fig:twobeams_offset} shows the transfer functions of the CHRONOS 2.5-m option obtained from the simulations. The common-mode response $\bar{H}(\omega)$ exhibited the expected $1/f^2$ slope, consistent with Eq.~(\ref{eq:H_TF}). We parameterized its deviation from the analytical model by an additive factor $\mu$ and modified the model as
\begin{equation}
\label{eq:H_TF_mu}
H^\prime(x_{\rm i},y_{\rm i};\omega)\equiv (1+\mu)\frac{x_{\rm i}^2}{I\omega^2}.
\end{equation}
The best-fit value of $\mu$ in the 0.004--100~Hz range was 0.35. When both beams were injected simultaneously, the displacements at the two beams canceled out, resulting in a residual structure with a $1/f^4$ slope in the frequency range from 0.004~Hz to 0.4~Hz. 
The physical origin of this residual and its deviation from the simple model are discussed in Sec.~\ref{sect:discussion_section}.

\subsection{Beam position mismatch: $\alpha$}
\label{subsec:beam_mismatch}

In a realistic interferometric system, the actual beam positions may deviate slightly from their ideal design locations. Such deviations can cause effective leakage of the common-mode into the differential-mode response. 

It has been reported that the beam spot was at 17 mm away from the center of the end test mass at LIGO Livingston Observatory during its third observing run, but it could be controlled to within 1 mm from the center~\cite{beamspot_2019,alog_beamspot_llo_2024}. Given that the CHRONOS arm cavity is ten times shorter length than that of Advanced LIGO, we assume that a 1-mm beam-position accuracy is achievable without serious difficulty. 

To evaluate the magnitude of the common-mode leakage effect caused by beam position mismatch, we performed FEA simulations in which the beam position was systematically shifted. In particular, when the left beam was displaced outward by 1~mm, a small leakage of the nominal $1/f^2$ component appeared in the residual of the differential-mode transfer function above 0.1~Hz. We parameterize the magnitude of this leakage by $\nu$, and that of $1/f^4$ component by $\delta$. Thus, $\Delta H(\omega)$ can be approximated as
\begin{equation}
\label{eq:Delta_H_TF}
\Delta H^\prime(\omega)\equiv \left( \frac{2\alpha+\nu}{I\omega^2}+\frac{\delta}{I^2 \omega^4} \right)\bar{x}^2.
\end{equation}
The best-fit values of $\nu$ and $\delta$ were -0.0019 and 0.0022, respectively.

The physical meaning of this effect and its implications for interferometer 
performance will be discussed in more detail in 
Sec.~\ref{sect:discussion_section}.

\subsection{Sensing function of the speed meter: $C(\omega)$}
\label{subsec:sensing}

The third ingredient required for the sensitivity evaluation is the sensing function \(C(\omega)\), which characterizes how the 
speed-meter readout converts the optical quadratures into the measured signal. 
In the CHRONOS configuration, \(C(\omega)\) depends on the optical layout 
(including the power- and signal-recycling cavities) as well as the chosen 
homodyne readout angle.  

\begin{figure}[t]
\begin{center}
\includegraphics[width=12.0cm]{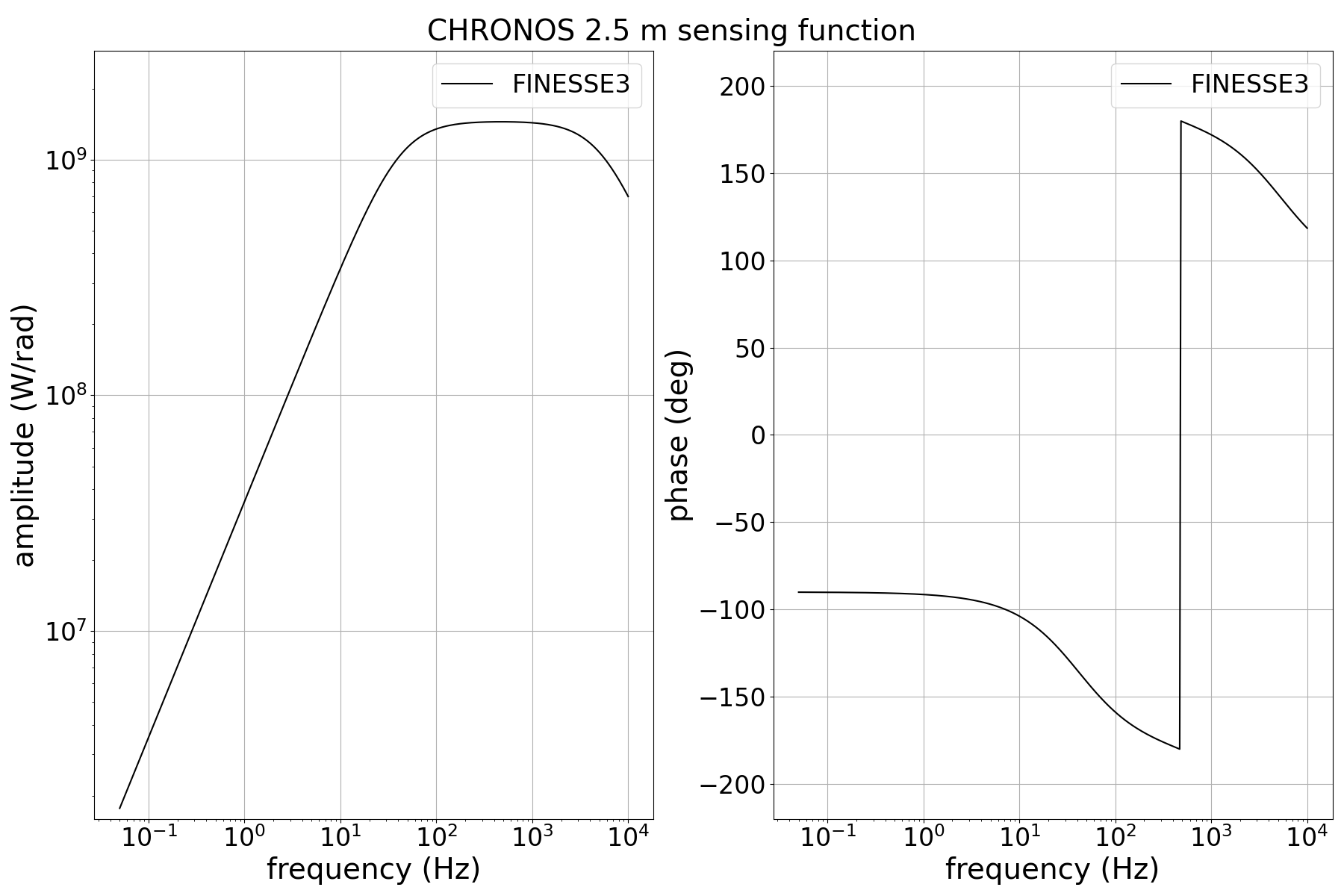}
\end{center}
\caption[]
{ \label{fig:sensing}
Sensing function of CHRONOS calculated by \textsc{FINESSE3}. }
\end{figure}

In this work, the sensing function \(C(\omega)\) was evaluated using 
\textsc{FINESSE3}, assuming the full CHRONOS optical configuration~\cite{finesse_software,CHRONOS}. Figure~\ref{fig:sensing} compares the resulting frequency dependence for the three options of the detector's scale. 
As shown, the functional shape varies systematically with the arm length, 
reflecting the trade-off between low-frequency response and bandwidth.  

%% file: 4_evaluation.tex
\section{Evaluation based on CHRONOS design\label{sect:evaluation_section}}

We calculated the predicted intensity-noise curve of CHRONOS based on Eq.~(\ref{eq:h_total_spectrum}). The frequency-independent parameters summarized in Table~\ref{tab:parameters} were used~\cite{CHRONOS}. As for the transfer functions $\Delta H(\omega)$ and $\bar{H}(\omega)$, we used the simulated curves obtained in Sec.~\ref{subsec:H} for the case in which the left beam was shifted outward by 1 mm (red and green curves in Fig.~\ref{fig:twobeams_offset}). With this beam offset, $\alpha$ is around 0.002 at a distance of 50 cm from the bar center. The mismatch of the two balanced-homodyne detection paths, $\gamma$, is typically 1\%. The sensing function was the one we calculated in Sec.~\ref{subsec:sensing}.

\begin{table}[tbp]
\begin{center}
\begin{tabular}{c|c|c}
\hline
Parameter (Unit) & Symbol & Value \\
\hline\hline
DC input power of laser (W) & $P_{\rm in}$ & 1 \\
Cavity arm power of laser (W) & $P_{\rm arm}$ & 444 \\
Beam position to bar center (mm) & $\bar{x}$ & 537.3 \\
Power fraction of local oscillator (\%) & $\beta$ & 0.1 \\
Beam position mismatch (\%) & $\alpha$ & 0.19 \\
Homodyne imbalance (\%) & $\gamma$ & 1 \\
Relative intensity noise (dB${\rm Hz}^{-1/2}$) & ${\rm RIN}(\omega)$ & -151 flat \\
\hline
\end{tabular}
\end{center}
\caption{Frequency-independent parameters used for evaluating intensity noise of the CHRONOS 2.5-m option. 
The values of $\alpha$, $\beta$, and $\gamma$ are assumptions, and the relative intensity noise is based on Advanced LIGO~\cite{ligo_intensity}. 
All other values are design parameters of the three arm-cavity options~\cite{CHRONOS}.}
\label{tab:parameters}
\end{table}


The ratio between the local-oscillator beam power and the main laser power, denoted by $\beta$, was chosen so that the power at each PD was 0.5~mW. Since the typical saturation power of an InGaAs PD is 1~mW, the local-oscillator beam power had a margin of one-half.

It has been reported that an intensity stabilization system with two feedback loops, the same structure as of that of Advanced LIGO during O4, achieved a RIN of $4\times 10^{-8}\,{\rm Hz}^{-1/2}$ (-148 dB$\,{\rm Hz}^{-1/2}$) at 1~Hz~\cite{ligo_intensity,ligo_O4}. After subtracting the intrinsic noise of the in-loop and out-of-loop PDs, which are designed to be equal, the true RIN becomes 3~dB lower. We assumed the same RIN, namely -151 dB$\,{\rm Hz}^{-1/2}$, at all frequencies as an input for our calculation. 

\begin{figure}
\begin{center}
\begin{tabular}{c}
\includegraphics[width=12.0cm]{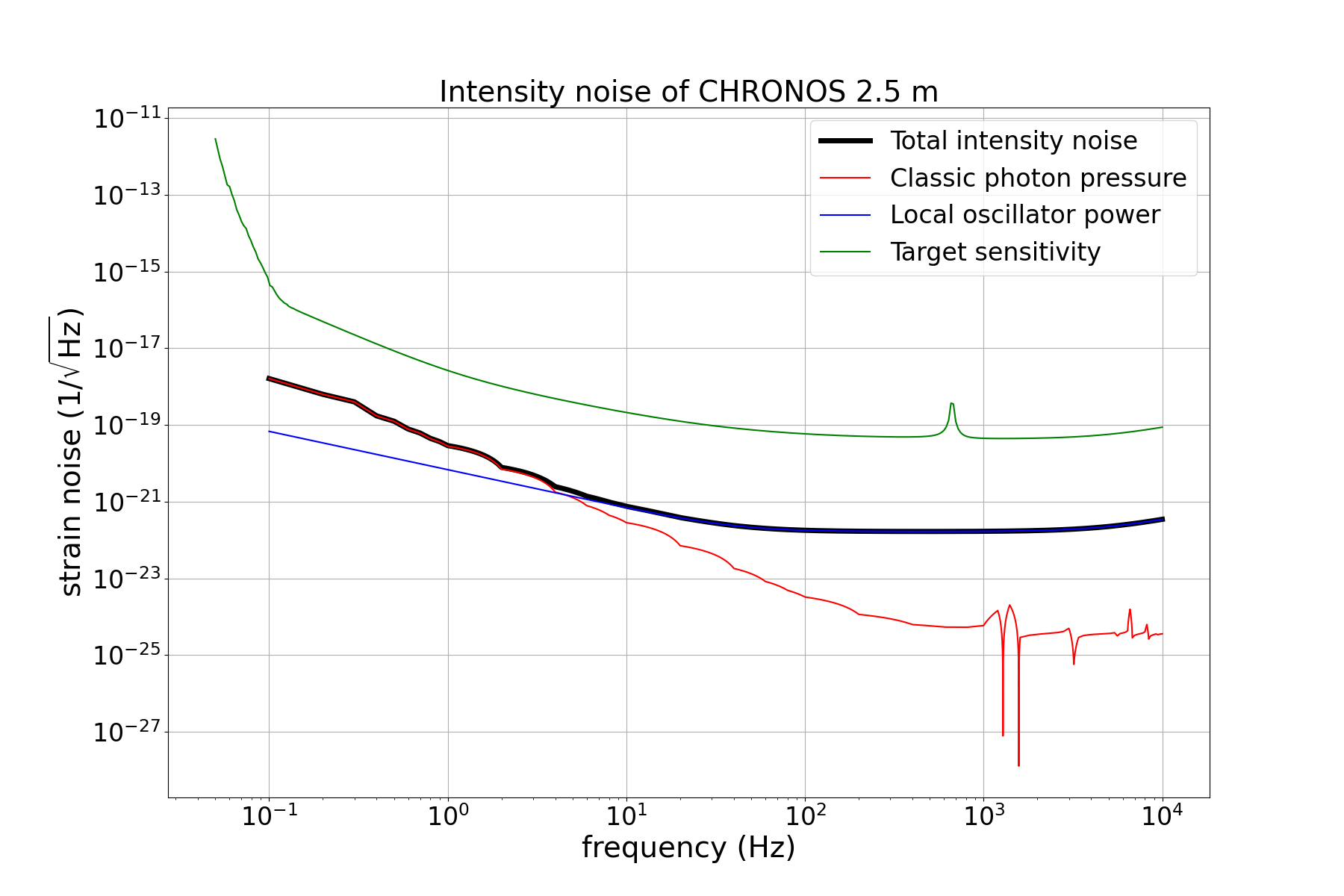}
\end{tabular}
\end{center}
\caption[]
{ \label{fig:chronos_intensity}
Intensity noise of the CHRONOS 2.5-m option and its noise budget in the unit of spacetime strain. ({\it Solid black}) Total intensity noise; ({\it Solid red}) Term of classic photon pressure which pushes the bar; ({\it Solid blue}) Term of local oscillator's beam power; ({\it Solid green}) Target sensitivity~\cite{CHRONOS}.
}
\end{figure}

Given all these parameters and spectra, the intensity noise of the 2.5-m option in terms of spacetime strain was calculated as Fig.~\ref{fig:chronos_intensity}. It predicted $2.9\times 10^{-20}\,{\rm Hz}^{-1/2}$ at 1 Hz.

%% file: 5_discussion.tex
\section{Discussions\label{sect:discussion_section}}

The projected intensity noise of the CHRONOS 2.5-m option is sufficiently smaller than the target sensitivity required for IMBH detection. The dominant coupling with RIN below 10 Hz is the torsion-bar response with a mismatch of the beam position.

Compared with a Michelson interferometer, the torque-cancellation effect suppresses the arm-length variation by a factor of 100. The variation in differential arm length per unit RIN per unit arm power was calculated to be $2.2\times 10^{-15}$~m$\cdot {\rm Hz}^{1/2}$/W based on Eq.~(\ref{eq:theta_bar}). In the case of Advanced LIGO, the differential arm length per RIN was on the order of $1\times 10^{-8}$~m$\cdot {\rm Hz}^{1/2}$, which corresponds to the order of $1\times 10^{-14}$~m$\cdot {\rm Hz}^{1/2}$/W after normalizing by an arm power of 800~kW~\cite{ligo_intensity_model_part3}. This suppression is a consequence of the cancellation of the $1/f^2$ component.

The frequency dependence of the local-oscillator component is determined by the sensing function. The sensing function of a speed meter has a slope proportional to $f$ at low frequencies, while that of a dual-recycling Fabry--P$\acute{e}$rot Michelson interferometer is proportional to $f^2$~\cite{ligo_sensing}. This suppresses the increase of the local-oscillator component in terms of intensity noise, and hence keeps it subdominant at frequencies below 4~Hz.

We have neglected DARM offset, cavity imbalance, and sideband transmission at the OMC in this paper. All these components depend on mirror properties. The high reflectivity of the CHRONOS mirrors, in particular, makes the cavity reflectivity sensitive to imbalance. This can couple with variations of reflectivity and path length of the triangular cavities on the left and right sides of the bar. However, these effects remain subdominant because they are suppressed by the inverse sensing function $C_{\rm modeled}^{-1}$ during the reconstruction phase, as long as they do not couple to the rotation of the bar. In addition, the OMC is designed to suppress sideband transmission to -40 dB~\cite{CHRONOS_optics}. 

The intensity-noise models of Advanced LIGO and KAGRA include asymmetries in reflectivity, cavity-pole frequency, cavity gain, and mirror mass as sources of arm-power fluctuation between the two cavities, and therefore as causes of imbalance in photon pressure~\cite{ligo_intensity_model_part3,kagra_intensity}. Our photon-pressure formula corresponds to a modification of the mirror-mass term in their model. The other three terms remain compatible with Advanced LIGO when they are normalized by arm power and RIN. Although they are not a specific to a torsion bar, precise fabrication mirror coatings is essential to suppress them. 

The transfer function of bar response was approximated by a $1/f^2$ dependence in Eq.~(\ref{eq:H_TF}). By contrast, Fig.~\ref{fig:twobeams_offset} showed some discrepancies from the analytical model. We parameterized these discrepancies by $\mu$, $\nu$, $\delta$ in Sec.~\ref{sect:offset_section}. 

The parameter $\mu$, which parametrizes a modification to the mean transfer function of the left and right beams, which is denoted by $\bar{H}(\omega)$ in Eq.~(\ref{eq:H_delta_bar}), can be interpreted as an additional contribution from translational motion. It may primarily arise from local deformation and translational motion. The parameter $\nu$, which was added to the $1/f^2$ leakage into the differential transfer function $\Delta H(\omega)$, may arise from asymmetry of the finite size of mesh. The parameter $\delta$, which characterizes the magnitude of the $1/f^4$ component, may also arise from the mesh asymmetry.

Using the approximate expressions in Eq.~(\ref{eq:H_TF_mu}) and Eq.~(\ref{eq:Delta_H_TF}), the intensity-noise formula Eq.~(\ref{eq:h_total_spectrum}) can be parameterized as
\begin{equation}
\label{eq:h_total_spectrum_approx}
\begin{array}{l}
S_h^\prime(\omega)\simeq {\rm RIN}(\omega) \\
\cdot \sqrt{ \frac{3P_{\rm arm}^2 \bar{x}^2} {2c^2 I\omega^2} \left| (1-\mu)\alpha+\nu+\frac{\delta}{I\omega^2} \right|^2+\left| C_{\rm modeled}^{-1}(\omega) \right|^2 P_{\rm in}^2 \beta^2\gamma^2}.
\end{array}
\end{equation}

\begin{figure*}
\begin{center}
\begin{tabular}{c}
\includegraphics[width=18.0cm]{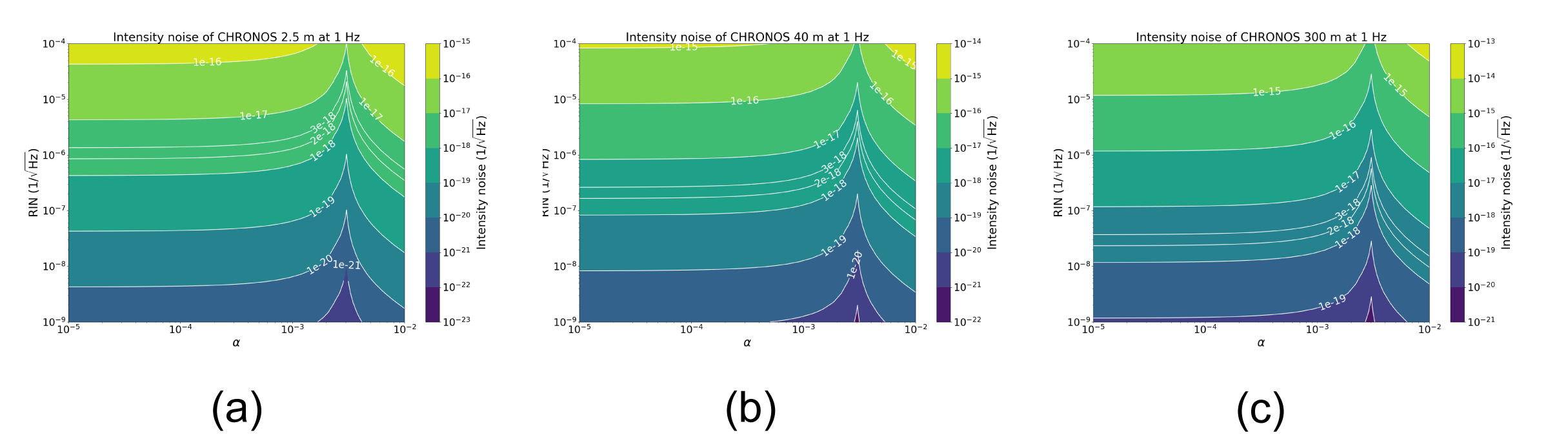}
\end{tabular}
\end{center}
\caption[width=18.0cm]
{ \label{fig:chronos_intensity_heatmap}
Intensity noise of CHRONOS calculated by the simplified analytical formula with the beam-position mismatch parameter $\alpha$ and RIN. (a) 2.5~m; (b) 40~m; (c) 300~m.
}
\end{figure*}

Figure~\ref{fig:chronos_intensity_heatmap} shows the intensity noise of CHRONOS at 1~Hz calculated from Eq.~(\ref{eq:h_total_spectrum_approx}) as a map over $\alpha$ and RIN. We assumed $\mu=0.35$, $\nu=-0.0019$, and $\delta=0.0022$. This yields the requirements on $\alpha$ and RIN needed to meet the target sensitivity. The target sensitivities of the 2.5-m, 40-m, 300-m options are $3\times 10^{-18} \, {\rm Hz}^{-1/2}$, $2\times 10^{-18} \, {\rm Hz}^{-1/2}$, and $1\times 10^{-18} \, {\rm Hz}^{-1/2}$, respectively. Considering that it is difficult to reduce $\alpha$ below 0.001 (corresponding to an alignment accuracy of 0.5~mm), this sets the following requirement for RIN: $2\times 10^{-6} \, {\rm Hz}^{-1/2}$ (-114~dB${\rm Hz}^{-1/2}$), $2\times 10^{-7} \, {\rm Hz}^{-1/2}$ (-134~dB${\rm Hz}^{-1/2}$), and $1\times 10^{-8} \, {\rm Hz}^{-1/2}$ (-160~dB${\rm Hz}^{-1/2}$). The RIN of -160~dB${\rm Hz}^{-1/2}$ at 1~Hz is at the same level as the DECIGO requirement~\cite{decigo_intensity}.

Although it is outside the science band of CHRONOS, we comment on the flat regions of $H(\omega)$. The flat region with several small peaks is dominated by bulk-deformation modes. The first mode around 700~Hz is not seen because the direction of deformation is perpendicular to the beam axis. The higher modes appear at the frequencies expected from the modal analysis. This is the first study to include bulk deformation in a sensitivity calculation of a GW detector, whereas previous studies considered it only in systematic-error estimation~\cite{ligo_pcal,kagra_pcal,bulk_document}.

The other flat region at low frequency has a knee at 0.004~Hz. It is not related to the resonant frequency of the yaw rotation because this simulation did not include the suspension fiber. This flat region and the $1/f^4$ slope also cannot be explained by an expansion of the neglected $\omega_0 \omega/Q$ term in Eq.~(\ref{eq:theta_GW}), because they are $10^8$ times larger than $\omega_0 \omega/Q$. One possible origin is a weak-spring assumption in the simulation, and another is coupling between local deformation and asymmetry of mesh. It is typical that FEA simulations assume virtual weak springs attached to the object to confine the center-of-mass motion. Second-order springs can explain the $1/f^4$ slope. The fitting parameter $\delta$ in Eq.~(\ref{eq:Delta_H_TF}) includes this simulation-originated effect. The residual in an actual experiment is therefore expected to be smaller than that of the simulation.

%% file: 6_conclusion.tex
\section{Conclusion\label{sect:conclusion_section}}

We proposed an intensity-noise model for a GW detector employing a Sagnac speed meter with a torsion bar. Our intensity-noise model consists of photon pressure on the bar and the power of the local-oscillator beam at the balanced-homodyne detection stage as carriers of laser-intensity fluctuations. Various configurations of power- and signal-recycling cavities can be represented by the sensing function without changing the overall functional form of the model. We evaluated the intensity noise of CHRONOS based on our model and obtained $2.9\times 10^{-20}\,{\rm Hz}^{-1/2}$ for the 2.5-m arm cavity option. The intensity noises was sufficiently below the projected sensitivity designed for the detection of GWs from $\mathcal{O}(10^4)\,M_\odot$ binary IMBH at 1~Hz. The required RIN to meet the target sensitivity of the 2.5~m option was -114 dB${\rm Hz}^{-1/2}$, and the requirements for the longer-arm options can be satisfied by realistic improvements in RIN.

To evaluate the CHRONOS intensity noise, we performed FEA to include bulk deformation of the bar in addition to rotation. We showed that the intensity noise is limited at high frequencies above 1~kHz due to bulk deformation modes. It is a new effect that has been taken into account in the sensitivity of GW detector for the first time.

A Sagnac speed meter with a torsion bar is expected to be a powerful tool for exploring low-frequency GWs and the rich astrophysical and high-energy physics imprinted in them. Our intensity-noise model will be essential to realizing this novel technique.